\documentclass[12pt]{article}
\usepackage[dvips]{graphicx}
\usepackage{axodraw}
\def\jnfont{\rm}
\def\AP#1,{{\jnfont Ann.\ Phys.\ (N.Y.)} {\bf #1},}
\def\EPJC#1,{{\jnfont Euro.\ Phys.\ J.\ C}\ {\bf #1},}
\def\JETPL#1,{{\jnfont JETP Lett.}\ {\bf #1},}
\def\NPB#1,{{\jnfont Nucl.\ Phys.}\ {\bf B#1},}
\def\PL#1,{{\jnfont Phys.\ Lett.}\ {\bf #1},}
\def\PLB#1,{{\jnfont Phys.\ Lett.\ B}~{\bf #1},}
\def\PRD#1,{{\jnfont Phys.\ Rev.\ D}~{\bf #1},}
\def\PRL#1,{{\jnfont Phys.\ Rev.\ Lett.}\ {\bf #1},}
\def\SJNP#1,{{\jnfont Sov.\ J. Nucl.\ Phys.}\ {\bf #1},}
\def\ZPC#1,{{\jnfont Z. Phys.\ C} {\bf #1},}
\let\to=\rightarrow
\def\B{\mathop{\rm B}}
\def\chargino{\tilde\chi_1^-}
\def\neutralino{\tilde\chi_1^0}
\def\se{\tilde e}
\def\stau{\tilde\tau}
\def\GeV{\ifmmode\,{\rm GeV}\else $\,\rm GeV$\fi}
\def\etal{{\it et al.}}
\begin{document}

\thispagestyle{empty}

\hbox to \hsize{\hfil\vtop{\hbox{TU-548}\hbox{hep-ph/9805246}\hbox{May 1997}}}
\vskip 15mm plus .5fil
\begin{center}
{\LARGE Probing the sgeneration spectrum\\[10pt] in chargino decays}\\
\vskip 10mm plus .5fil
{\large Ken-ichi Hikasa and Takaaki Nagano}\\
\vskip 7mm
{\it Department of Physics, Tohoku University\\[5pt] 
Aoba-ku, Sendai 980-8578, Japan}
\end{center}
\vskip 30mm plus .5fil

Some mechanisms of supersymmetry breaking result in splitting of the third 
generation sfermions from the other sfermions.   We show that the three-body 
decay branching ratio of the lighter chargino gives a sensitive probe of the 
sfermion mass spectrum if the chargino has a large gaugino component.

\vfil
\newpage

If supersymmetry is the Nature's choice for the solution of the naturalness 
problem, a plethora of new particles should exist just above the weak scale.  
Each known particle is accompanied by its superpartner.  
The pattern of the sparticle mass spectrum reflects the mechanism of 
supersymmetry breaking.  One of the interest lies in the generation structure 
of the sfermion masses.  Although the neutral kaon system gives a strong 
constraint~\cite{KKbar}   
that the first and second generation squarks should be nearly 
degenerate or very massive, the third generation sfermions are much less 
constrained.  Indeed, in the supergravity scenario of supersymmetry breaking, 
mass splitting of the third-generation and the other sfermions results from 
the renormalization group evolution of the masses between the 
unification scale and the weak scale, even if the sfermions 
have equal masses at the unification scale.  This splitting is due to the 
effect of the large Yukawa coupling of the top.  The bottom and tau sectors 
are also affected if the parameter $\tan\beta$ (the ratio of 
the two higgs vacuum expectation values) is large.
There are also arguments~\cite{heavysf} 
that first and second generation sfermions 
can be as heavy as 10 TeV without conflicting the naturalness problem, 
while the third generation sfermions have to be rather light.

Electron-positron colliders are best machines to search for colorless 
superparticles.  Much effort has been devoted to supersymmetry 
searches at LEP, each time its energy is raised~\cite{LEPsearches}.  
Linear colliders 
with higher energies are in the stage of planning, which would extend 
the scope of the searches.  If superparticle mass spectrum is widely 
distributed, we can hardly expect that all superparticles are found 
at one experiment.  Rather, first discovery would be confined to one 
or some of the lighter superparticles, heavier ones being 
beyond the reach of the machine.   It would be desirable if one can 
extract some information on the heavier superparticles by studying 
the properties of the lighter ones.  

There are two possibilities of doing this.  One is to exploit the 
supersymmetric relations between couplings such as the equality of 
gauge couplings and gaugino couplings.  These relations receive 
corrections from supersymmetry breaking which depend logarithmically 
on heavy superparticle masses~\cite{HN}.  
This effect (sometimes called the `super-oblique' correction) has been 
studied recently by several groups~\cite{NojiriF}--\cite{Randall}.  

Another possibility is to look for the effect of virtual sparticle 
exchanges.  Unlike the superoblique corrections, this effect vanishes if 
the superparticle mass is large (the decoupling theorem).  
We find, however, there are cases in which this effect is quantitatively 
important~\cite{Nagano}.

In this paper we present a test case of the latter possibility.  One of the 
likely first superparticles is the (lighter) chargino, which generally is a 
mixture of the wino and the charged higgsino.   Charginos can be 
produced in pairs in $e^+e^-$ collisions.  If the two body decay channels 
such as $\hbox{neutralino} +W$ are not open, the chargino decays to a 
neutralino and a fermion-antifermion pair.  The three-body decay is important 
in a substantial region of the parameter space.  In this region, 
the branching ratio of the modes depends on the masses of the intermediate 
sfermions.  The ratio of leptonic and hadronic decay rate has been 
discussed by Oshimo and Kizukuri~\cite{OK}. 
We concentrate on the leptonic decays in the present paper 
and study the dependence of the branching ratio on the slepton masses, 
allowing uneqaul selectron/smuon and stau masses.  The phenomenological 
consequences of light staus have been examined in the framework 
of the minimal supergravity scenario by Baer \etal~\cite{Baer}.

Determination of supersymmetry parameters using chargino pair production 
has been studied by several groups~\cite{TFMYO}--\cite{FStwo}.  It was 
demonstated in these works that the chargino-neutralino sector 
parameters $M_2$, $\mu$, $\tan\beta$ as well as the electron sneutrino 
mass can be deduced from chargino production cross sections and 
various distributions in $e^+e^-$ collisions.  
In this work, we assume that these parameters are well determined 
and concentrate on the effect of slepton masses.  
Feng and Strassler~\cite{FStwo} and one of us~\cite{Nagano} 
have also examined the ratio of leptonic and hadronic branching fractions.  


The chargino-neutralino sector in the minimal supersymmetric standard 
model can be described by three unknown parameters, if one assumes 
the grand unification of the gaugino masses.  
They are the SU(2) gaugino mass $M_2$, the higgsino mass parameter $\mu$, 
and $\tan\beta$.  The hypercharge U(1) gaugino mass $M_1$ is given by 
$M_1={5\over 3}M_2\tan^2\!\theta_W \simeq 0.5 M_2$.  


We fix the mass of the lighter chargino $\tilde\chi_1^\pm$ (hereafter 
the chargino) as one of the parameters.  The remaining two may be chosen 
as $M_2/\mu$ and $\tan\beta$.  The composition of $\chargino$ is 
mainly determined by $M_2/\mu$.  It is mostly wino if $M_2/\mu\ll1$ (gaugino 
region) and mostly higgsino 
$\chargino\simeq \widetilde H_{1L}^- + \widetilde H_{2R}^-$ 
if $M_2/\mu\gg1$ (Higgsino region).  
In the region $M_2/\mu\sim1$, $\tilde\chi_1^\pm$ contains 
substantial components of the both.  The mixing angles have slight dependence 
on $\tan\beta$ and its sign (or equivalently the sign of $\mu$).

In the gaugino region, the lightest neutralino $\tilde\chi_1^0$ 
is mostly the bino (hypercharge gaugino) and its mass is about a half of 
the chargino mass.  The second lightest neutralino $\tilde\chi_2^0$ is 
mostly the neutral wino (an SU(2) gaugino) and is nearly degenerate with 
the chargino.  In the higgsino region, the lightest neutralino is 
a higgsino state which almost degenerates with (but lighter than) the 
chargino.  The second lightest neutralino, which is also a higgsino state, 
lies just above the chargino.  

Now we turn to the decay modes of $\chargino$.   The final state must contain 
one superparticle in the minimal framework with $R$ parity conservation.  
We are interested in the case that the sfermions are heavier than the 
chargino and the decays $\chargino\to \tilde f + \bar f$ are not allowed.  
If the mass difference between $\chargino$ and $\neutralino$ is 
large enough, the chargino decays by emitting a $W$ boson 
$ \chargino \to \neutralino + W^- $
or by emitting a charged Higgs boson (which is usually heavier than 
the $W$).  If these decays are kinematically forbidden, the main decay 
modes are
\begin{eqnarray}
& &\chargino \to \neutralino + q_d + \bar q_u \nonumber\\
& &\chargino \to \neutralino + \ell + \bar\nu \label{eqldecay}
\end{eqnarray}
where $q_d$ ($q_u$) denotes a charge $-1/3$ (2/3) quark.  The two-body 
decays are forbidden if the chargino is lighter than $\sim 180$ GeV 
in the wino region, and in most of the parameter space in the higgsino 
region (see Fig.~1).  

Now we concentrate on the three body leptonic decay of the chargino 
(\ref{eqldecay}).  In the lowest order of the electroweak couplings it 
proceeds via the exchange of $W$, charged Higgs, slepton, or sneutrino.  
The Feynman graphs are shown in Fig.~2.  If the decay is 
dominated by the $W$ exchange, the branching ratios are determined 
by gauge universality
\begin{equation}
{B_\tau\over B_e} \equiv 
{\B(\chargino\to\neutralino\tau\nu)\over\B(\chargino\to\neutralino e\nu)} 
= {\Gamma(\chargino\to\neutralino\tau\nu)\over%
\Gamma(\chargino\to\neutralino e\nu)}
= 1 
\end{equation}
(the lepton masses are neglected).  
Even when the sfermion exchanges contribute, the ratio of the branching 
ratios remains unity if the sfermions of different generations have 
a common mass.  Deviation from this prediction thus signals the splitting 
of the third generation sleptons from the first/second generations.  
The $H^-$ exchange is negligible for the decay to electron or muon, but not 
necessarily so for the tau decay, especially when $\tan\beta$ is large.  
This gives additional flavor-dependent effect, but it turns out to be 
not so important.  

If the lepton mass and Yukawa coupling are neglected, only the wino component 
of $\chargino$ interacts and the right-hand component of 
the charged lepton and the right-handed sfermion totally decouple.  
The charged Higgs also does not contribute.  The decay rate in this 
case depends on one additional parameter other than those of the 
chargino-neutralino sector, which is the mass of the left-handed sleptons 
($\widetilde\nu$ and $\tilde\ell_L$ are split by the $D$ term effect 
proportional to $m_Z^2\cos2\beta$).  This applies to the selectron 
and smuon decays.  

For tauonic decay we have to retain the Yukawa coupling (though we neglect 
the kinematic tau mass), through which the right-handed tau/stau 
components and charged Higgs affect the amplitude.  We find that 
the dependence of the rate on these masses is weak, unless these particles 
are much lighter than the left-handed sleptons.  

We now show the results of the calculation of the $\tau/e$ ratio.  
We fix the chargino mass $m(\chargino)=150\GeV$, which is under 
the real $W$ threshold for any $M_2/\mu$.  The results are not sensitive 
to the chargino mass as long as we stay below the threshold.  The decay 
rates are calculated for several values of $M_2/\mu$ and $\tan\beta$, 
varying the slepton masses $m(\se_L)$ and $m(\stau_L)$.  

The right-handed sleptons are assumed to be degenerate with the 
left-handed sleptons of the same flavor.  The results change very slightly 
if the right-handed sleptons are half as heavy.  The left-right mixing 
is also neglected.  Although the mixing may be substantial for staus, 
the decay rate is essentially determined by the left-left element of the 
sfermion mass matrix in most cases.

In Fig.~3, we show the $\tau/e$ ratio for $\tan\beta=50$ for 
$M_2/\mu=0.1$, 0.4 and 1.  
The ratio drastically deviates from unity in the gaugino region, even when 
the slepton mass is much larger than the chargino mass.  
The ratio can be as large as ten in some parameter region, although 
the sfermions are much heavier than the $W$ boson.  
The reason of this behavior is that the bino, which comprise the largest 
part of the lightest neutralino has no gauge coupling.  The decay via 
virtual $W$ occurs either via the small wino component of $\neutralino$ 
or the higgsino component of both $\chargino$ and $\neutralino$.  

The behavior of the curves in Fig.~3(a) may be understood in the following 
way.   If the $W$ exchange is negligible, the branching ratio is controlled 
by the mass of the exchanged sleptons and we find
$$ B_\tau/B_e \simeq m_{\se}^4/m_{\stau}^4 \;.$$
This dependence is approximately realized in the lower left region of 
Fig.~3(a), in which both the sleptons are relatively light.  
In the lower-right region, the selectron exchange becomes smaller than 
the $W$ exchange, so the above behavior switches to 
$$ B_\tau/B_e \sim m_W^4/m_{\stau}^4 \;.$$
On the diagonal line $m_{\se}=m_{\stau}$, we expect $B_\tau/B_e=1$.  
Slight deviation of the equal branching curve from the diagonal line in 
Fig.~3(a) is due to the charged Higgs exchange (We have taken 
$m_H=250\GeV)$.

The importance of the Higgs exchange can be seen from Fig.~4, in which 
we plot the equal-BR lines $B_\tau/B_e=1$ for different Higgs masses.  
We can see that the charged Higgs has little effect 
if its mass is larger than, say, twice as the slepton mass.

In the mixed region $M_2/\mu\sim1$, the sfermion mass dependence 
is rather weak since the $W$ exchange gives dominant contribution 
(the couplings are similar and there is no mass suppression).  
In the higgsino region (not shown) the sfermion couplings are essentially 
the Yukawa couplings and the branching ratios are insensitive to the 
sfermion mass.

In Fig.~5, the ratio $B_\tau/B_e$ is plotted for $\tan\beta=4$ and $-50$.  
The 
deviation is larger for negative $\tan\beta$ (equivalently negative $\mu$).  
This comes from the fact that the higgsino component in $\chargino$ is 
smaller for negative $\tan\beta$.  The deviation is somewhat milder for 
smaller $\tan\beta$.

To summarize, we have demonstrated that the measurement of the branching 
ratios of the three-body chargino decays can give information on the mass 
of the exchanged sfermions.  The flavor nonuniversality of the leptonic 
branching ratios implies the flavor nondegeneracy of sleptons unless 
the charged Higgs is light.  The effect is especially prominent when the 
chargino is wino-like and stronger for larger $\tan\beta$.  The flavor 
dependence of slepton masses, which may cast a light to the mechanism of 
supersymmetry breaking, may be thus probed even when the sleptons cannot be 
directly produced.   

The measurement of the branching ratios does not fix both $m(\se)$ 
and $m(\stau)$.  If selectron (therefore electron sneutrino) is not too heavy, 
the mass $m(\widetilde\nu_e)$ may be extracted from the differential 
chargino pair production cross section~\cite{FPMT}.  If this is the case, 
determination of both slepton masses is possible.


\vspace{10mm}

This work is partly supported by the Grant-in-Aid for Scientific Research 
(No.~08640343) from the Japan Ministry of Education, Science, Sports, and 
Culture.  


\newpage
\pagestyle{empty}
\newcounter{fig}
\begin{list}{Fig.~\arabic{fig}.\enspace}{\usecounter{fig}}
\item
{Mass difference between $\chargino$ and $\neutralino$.  
The two decay $\chargino\to\neutralino W^-$ is allowed only in the shaded 
region.}
\item
{Feynman diagrams for $\chargino\to\neutralino\ell^-\bar\nu$.}
\item
{The ratio $B_\tau/B_e$ for $m(\chargino)=150\GeV$, 
$\tan\beta=50$, $m(H^-)=250\GeV$, with (a) $M_2/\mu=0.1$; (b) 0.4; (c) 1.}
\item
{The equal-BR lines $B_\tau/B_e=1.0$ for $m(\chargino)=150\GeV$, 
$M_2/\mu=0.1$, $\tan\beta=50$, with $m(H^-)=200$, 
250, 300, 400, 500, 1000$\GeV$.}
\item
{The ratio $B_\tau/B_e$ for $m(\chargino)=150\GeV$, 
$M_2/\mu=0.1$, $m(H^-)=250\GeV$, with (a) $\tan\beta=-50$, (b) 4.}
\end{list}

\begin{figure}[htbp]
  \vspace*{0cm}
  \hspace*{-2.5cm}
  \includegraphics[angle=-90]{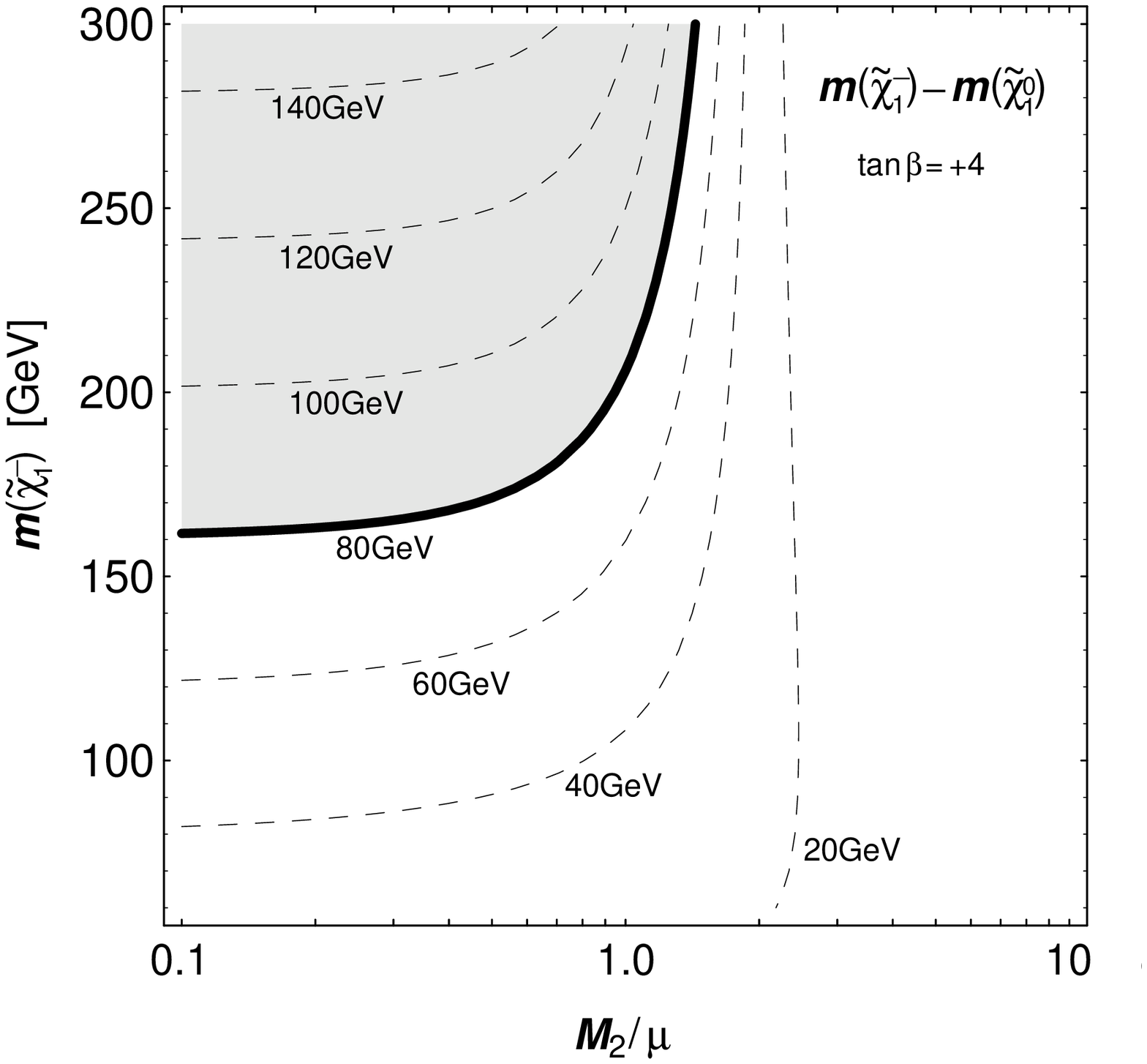}
  \vspace*{3cm}
  \begin{center}
    Figure~1
  \end{center}
\end{figure}

\begin{figure}[htbp]
\begin{center}
\begin{picture}(100,120)(-50,-60)
\Large
  \ArrowLine(0,-40)(0,0)
  \ArrowLine(0,0)(-40,40)
  \ArrowLine(40,40)(20,20)
  \ArrowLine(20,20)(0,40)
  \Photon(0,0)(20,20){3}{3.5}
  \Text(0,-44)[t]{$\tilde{\chi}^-_1$}
  \Text(-40,44)[b]{$\tilde{\chi}^0_1$}
  \Text(0,44)[b]{$\ell^-$}
  \Text(40,44)[b]{$\bar{\nu}$}
  \Text(15,5)[lt]{$W^-$}
\end{picture}~
\begin{picture}(100,120)(-50,-60)
\Large
  \ArrowLine(0,-40)(0,0)
  \ArrowLine(0,0)(-40,40)
  \ArrowLine(40,40)(20,20)
  \ArrowLine(20,20)(0,40)
  \DashLine(0,0)(20,20){4}
  \Text(0,-44)[t]{$\tilde{\chi}^-_1$}
  \Text(-40,44)[b]{$\tilde{\chi}^0_1$}
  \Text(0,44)[b]{$\ell^-$}
  \Text(40,44)[b]{$\bar{\nu}$}
  \Text(15,5)[lt]{$H^-$}
\end{picture}~
\begin{picture}(100,120)(-50,-60)
\Large
  \ArrowLine(0,-40)(0,0)
  \Line(0,0)(0,16)
  \ArrowLine(0,16)(0,40)
  \ArrowLine(-40,40)(-20,10)
  \Line(-20,10)(-6,17)
  \ArrowLine(6,23)(40,40)
  \DashLine(0,0)(-20,10){4}
  \Text(0,-44)[t]{$\tilde{\chi}^-_1$}
  \Text(-40,44)[b]{$\tilde{\chi}^0_1$}
  \Text(0,44)[b]{$\ell^-$}
  \Text(40,44)[b]{$\bar{\nu}$}
  \Text(-15,5)[rt]{$\tilde{\nu}_{L}$}
\end{picture}~
\begin{picture}(100,120)(-50,-60)
\Large
  \ArrowLine(0,-40)(0,0)
  \ArrowLine(0,0)(40,40)
  \ArrowLine(-40,40)(-20,20)
  \ArrowLine(-20,20)(0,40)
  \DashLine(0,0)(-20,20){4}
  \Text(0,-44)[t]{$\tilde{\chi}^-_1$}
  \Text(-40,44)[b]{$\tilde{\chi}^0_1$}
  \Text(0,44)[b]{$\ell^-$}
  \Text(40,44)[b]{$\bar{\nu}$}
  \Text(-15,5)[rt]{$\tilde{\ell}_{L/R}$}
\end{picture}
%
\end{center}
  \vspace*{3cm}
  \begin{center}
    Figure~2
  \end{center}
\end{figure}

\begin{figure}[htbp]
  \vspace*{0cm}
  \hspace*{-2.6cm}
  \includegraphics[angle=-90]{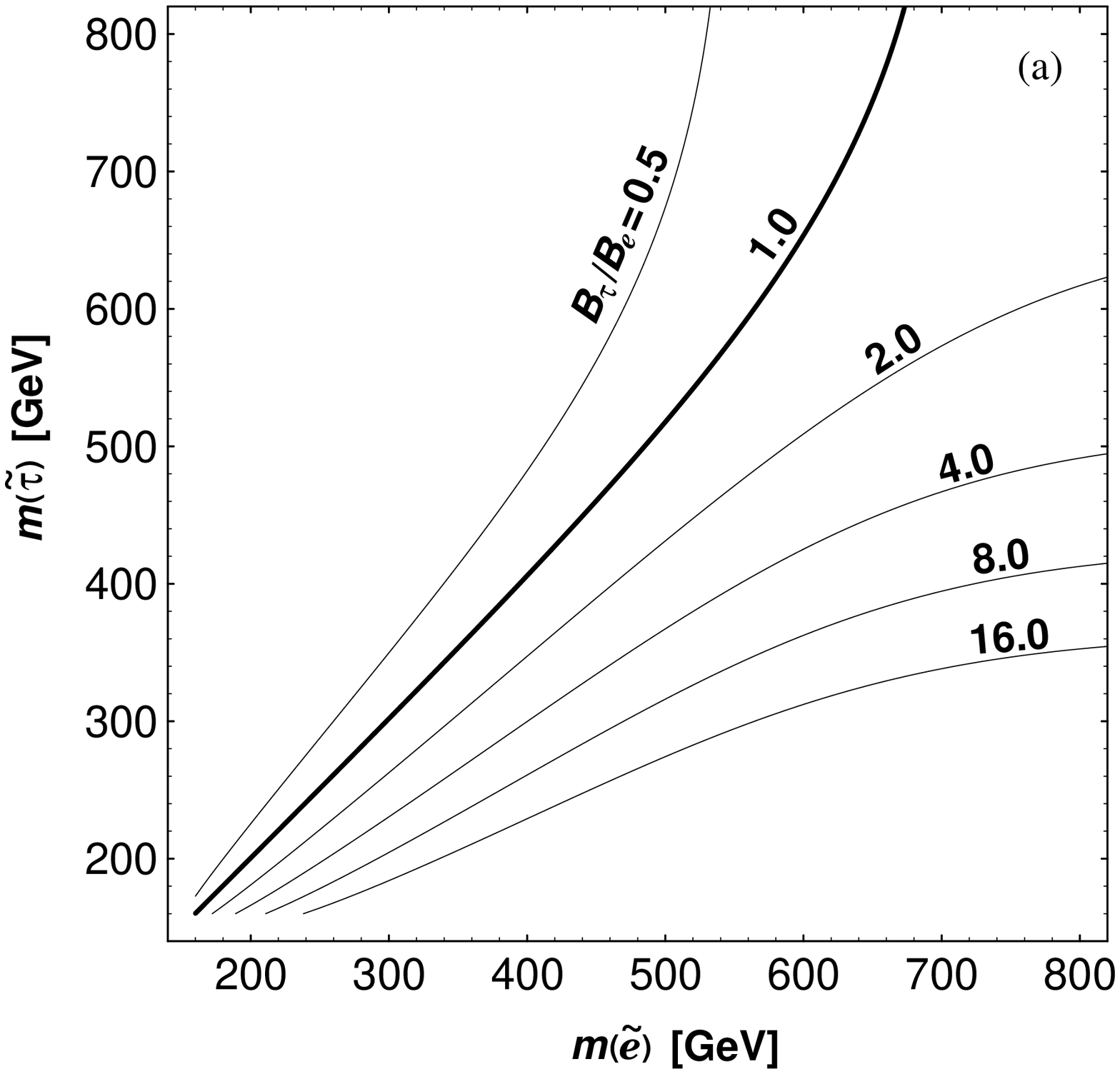}
  \vspace*{2cm}
  \begin{center}
    Figure~3(a)
  \end{center}
\end{figure}

\begin{figure}[htbp]
  \vspace*{0cm}
  \hspace*{-2.6cm}
  \includegraphics[angle=-90]{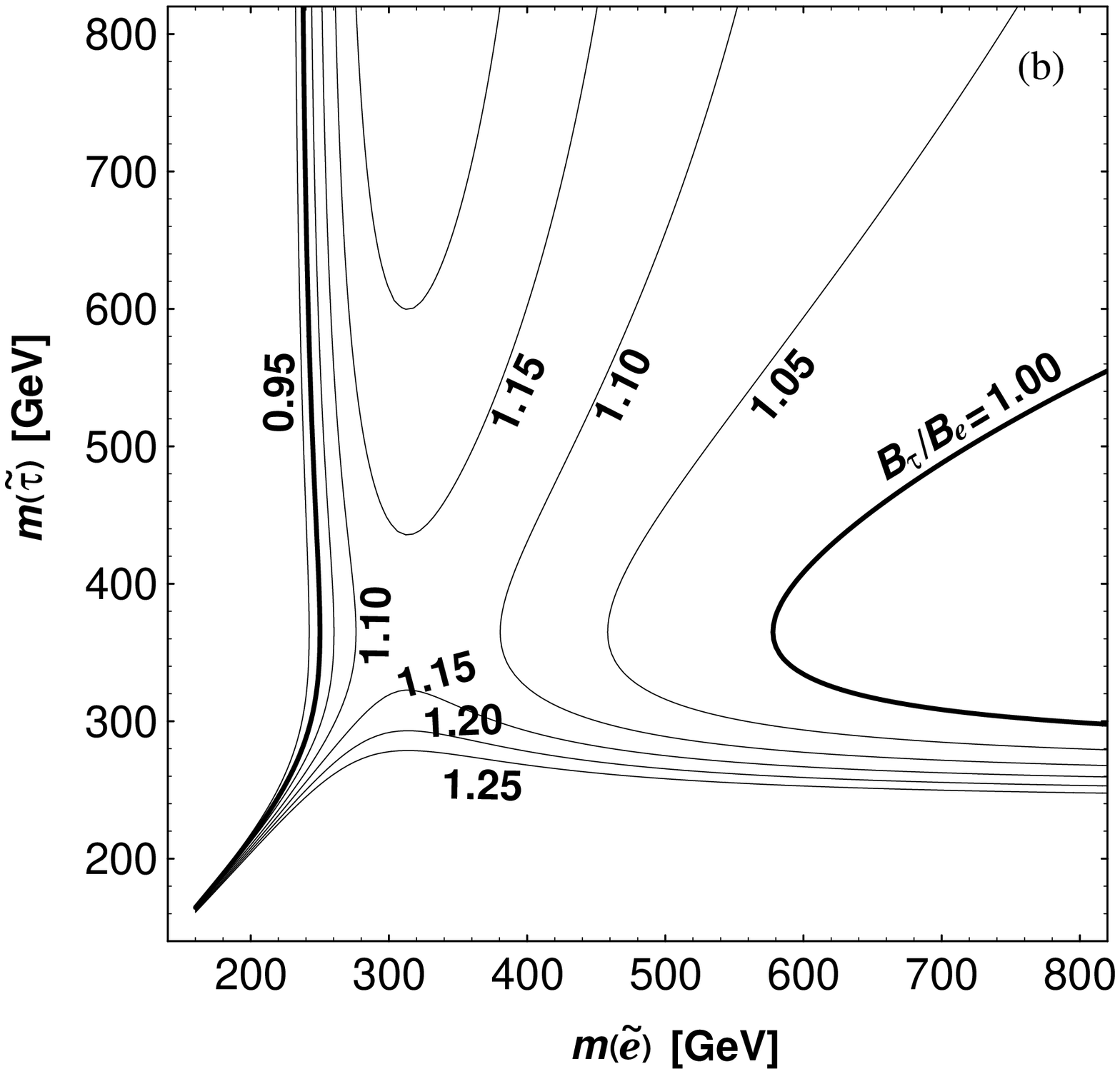}
  \vspace*{2cm}
  \begin{center}
    Figure~3(b)
  \end{center}
\end{figure}

\begin{figure}[htbp]
  \vspace*{0cm}
  \hspace*{-2.6cm}
  \includegraphics[angle=-90]{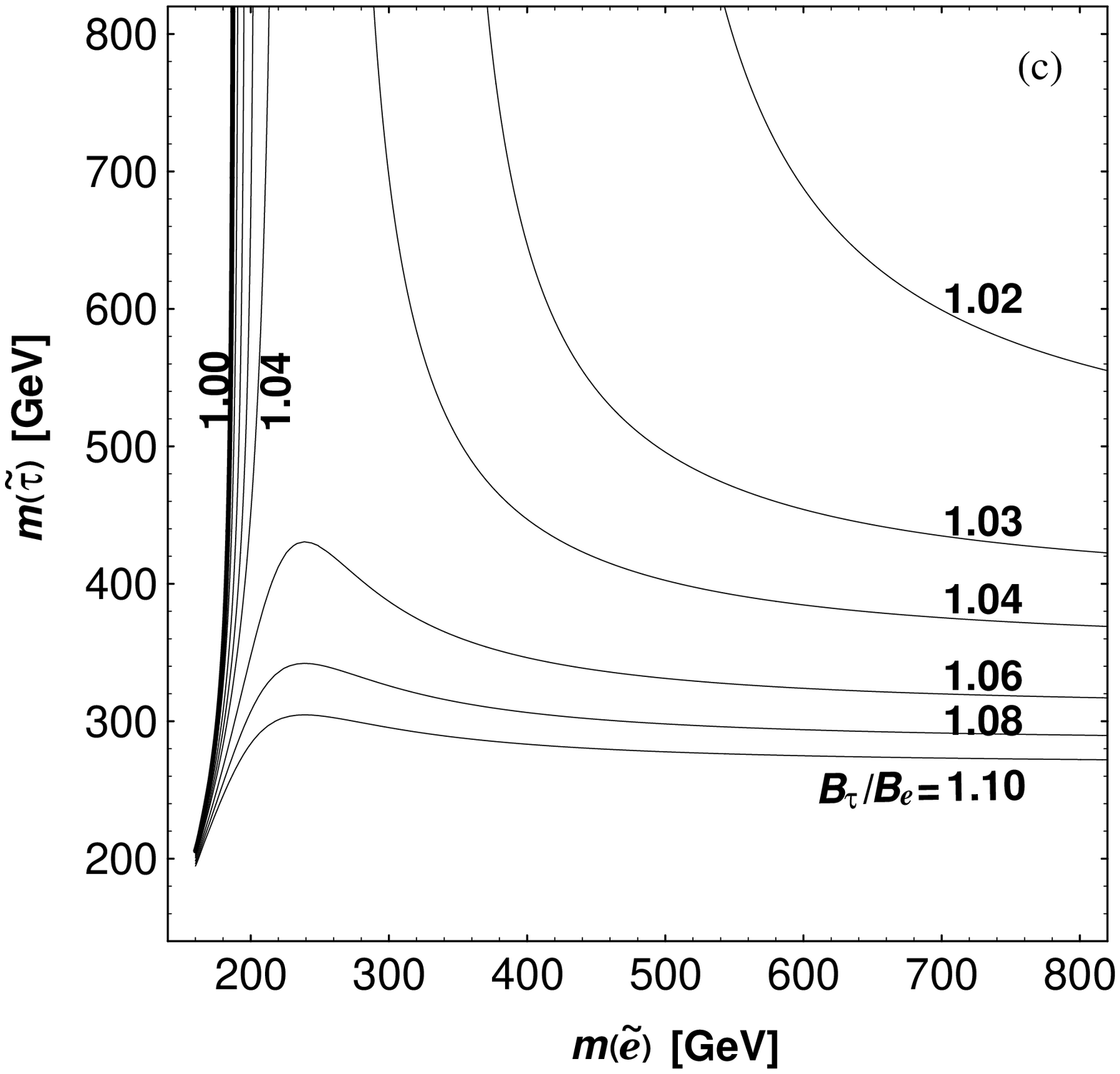}
  \vspace*{2cm}
  \begin{center}
    Figure~3(c)
  \end{center}
\end{figure}

\begin{figure}[htbp]
  \vspace*{0cm}
  \hspace*{-2.6cm}
  \includegraphics[angle=-90]{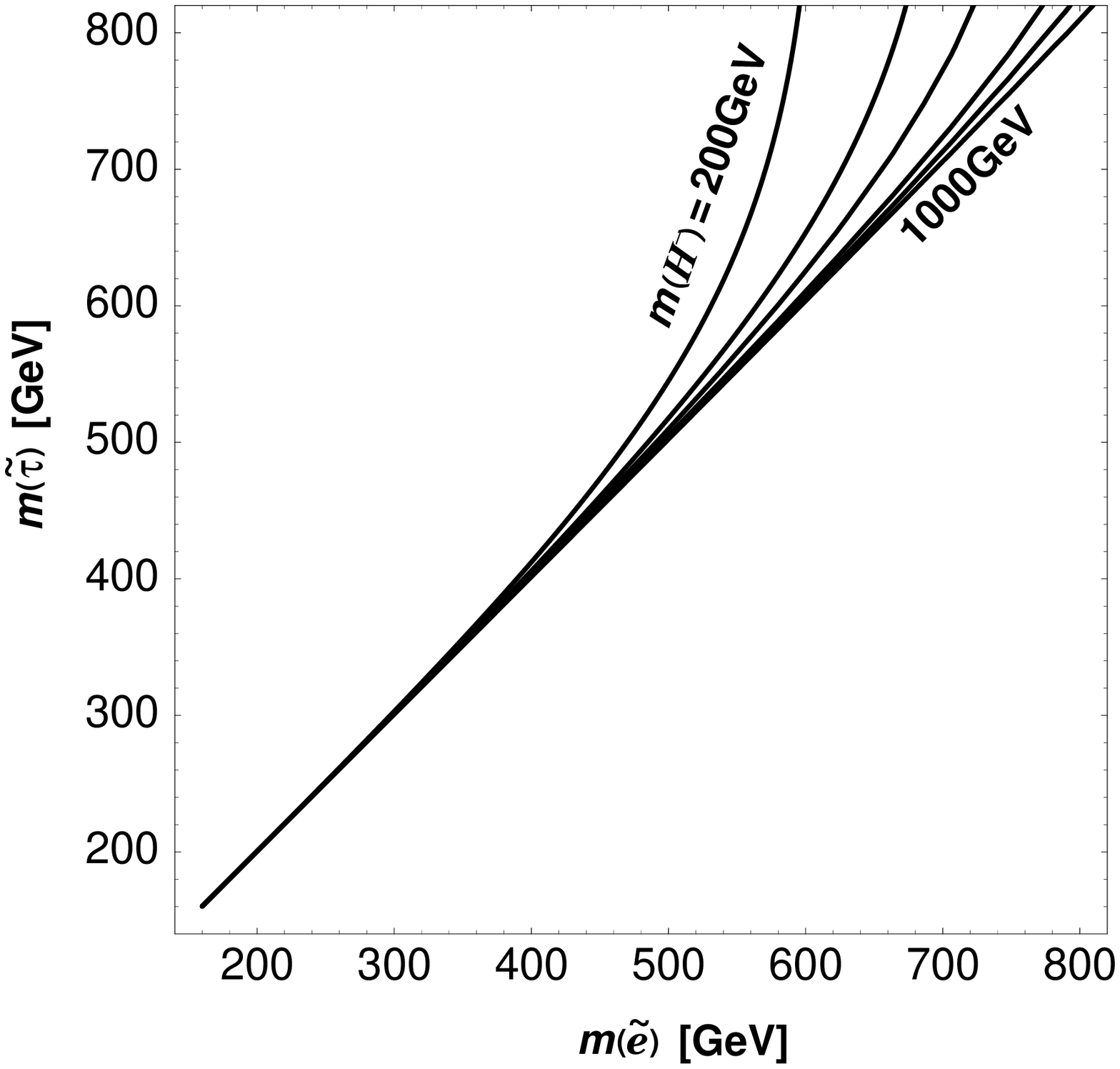}
  \vspace*{2cm}
  \begin{center}
    Figure~4
  \end{center}
\end{figure}

\begin{figure}[htbp]
  \vspace*{0cm}
  \hspace*{-2.6cm}
  \includegraphics[angle=-90]{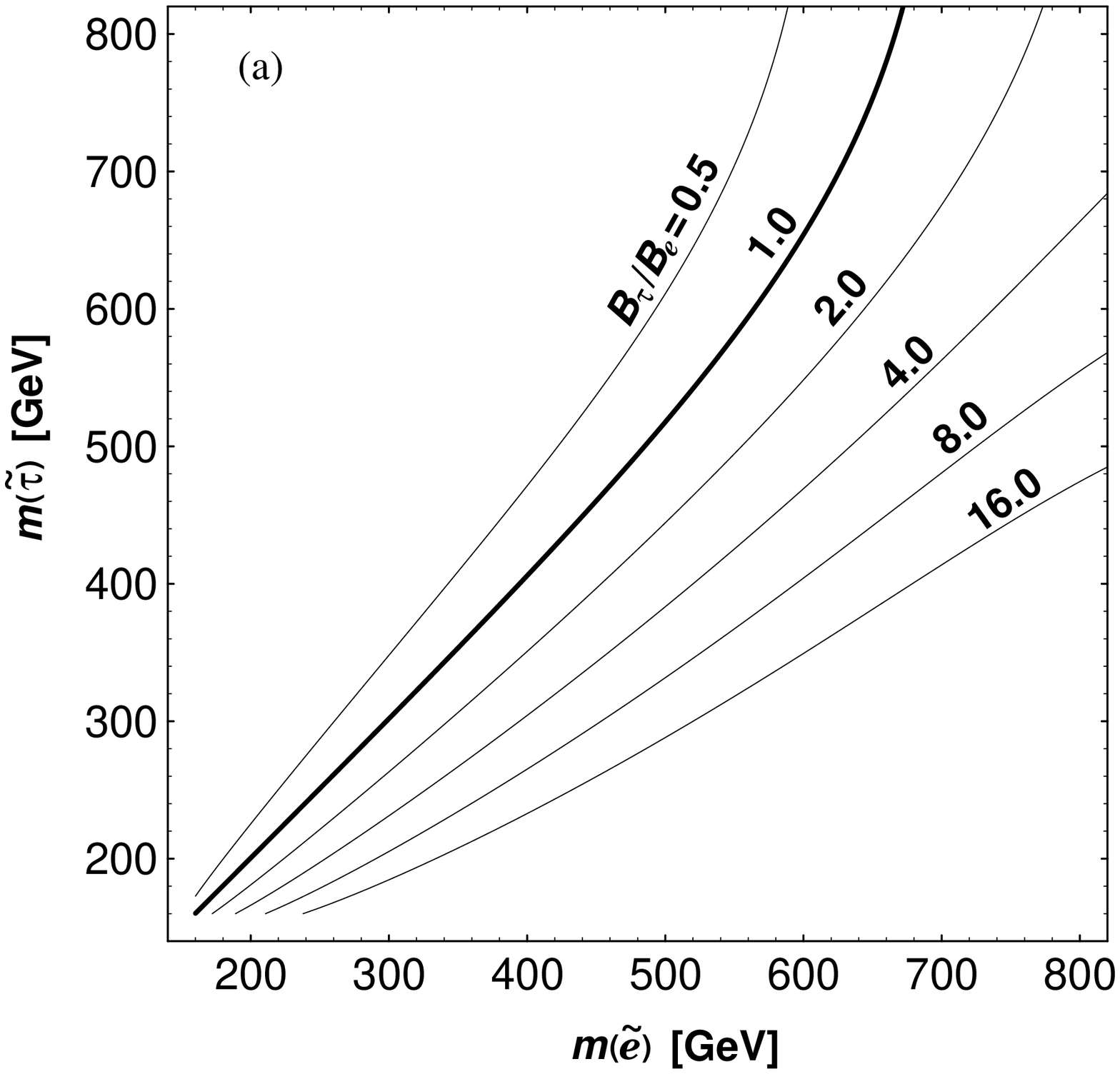}
  \vspace*{2cm}
  \begin{center}
    Figure~5(a)
  \end{center}
\end{figure}

\begin{figure}[htbp]
  \vspace*{0cm}
  \hspace*{-2.6cm}
  \includegraphics[angle=-90]{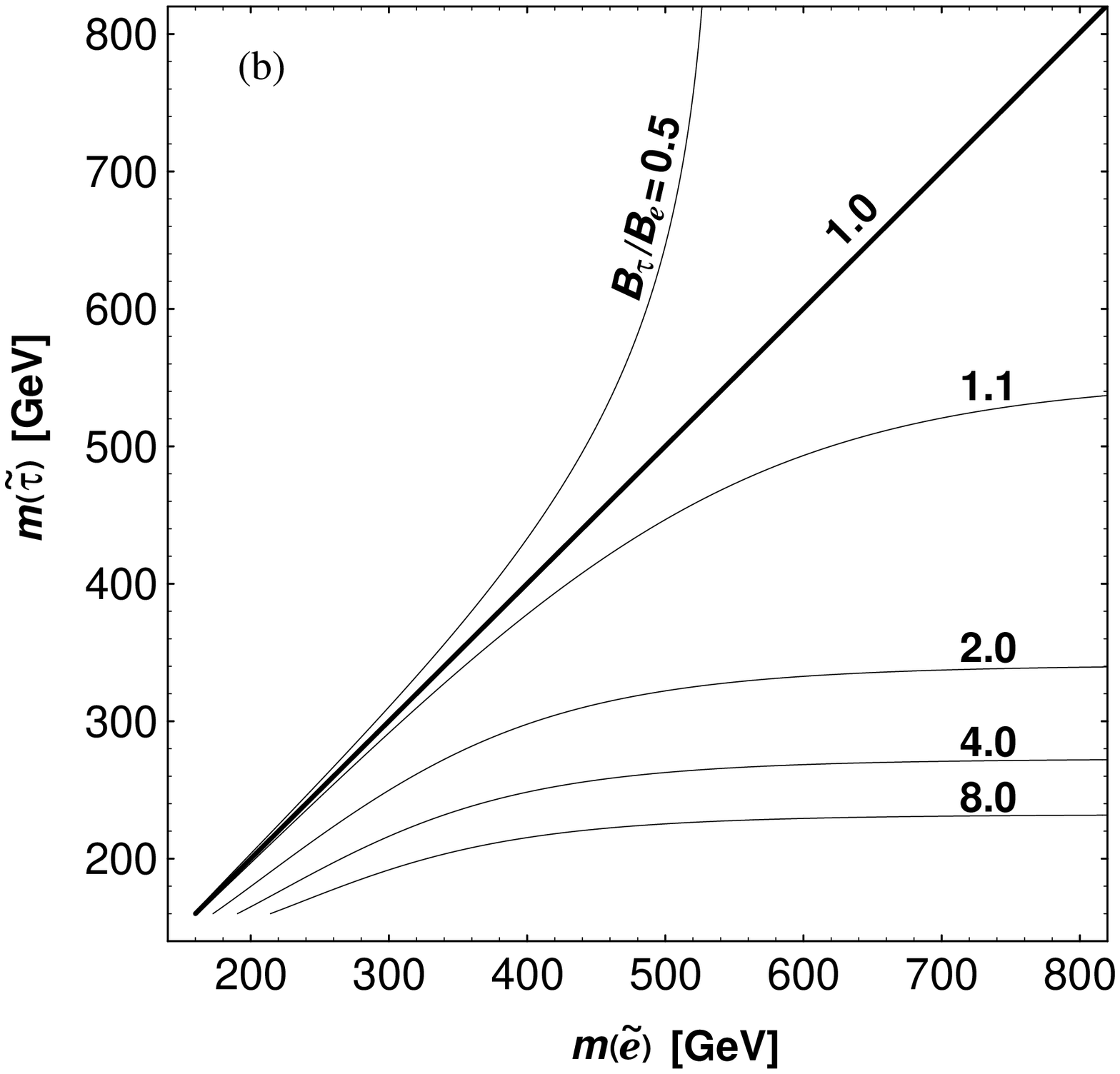}
  \vspace*{2cm}
  \begin{center}
    Figure~5(b)
  \end{center}
\end{figure}

\end{document}